\documentclass[12pt]{article}
\usepackage{graphics,epsfig}

\newcounter{saveeqn}

\begin{document}

\begin{titlepage}\hfill HD-THEP-09-14\\[5 ex]

\begin{center}{\Large \bf Lowest order covariant  averaging of a 
perturbed  
metric and of the Einstein tensor II} \\[5 ex]

{\bf Dieter Gromes}\\[3 ex]Institut f\"ur
Theoretische Physik der Universit\"at Heidelberg\\ Philosophenweg 16,
D-69120 Heidelberg \\ E - mail: d.gromes@thphys.uni-heidelberg.de \\
 \end{center} \vspace{2cm}

{\bf Abstract:} We generalize and simplify an earlier approach. In three
dimensions we present the most general  averaging formula in lowest order
which respects the requirements of covariance. It involves a bitensor,
made up of a basis of six tensors, 
and contains three arbitrary functions, which are only restricted by their 
behavior near the origin.  The averaging formula can also be applied to 
the Einstein tensor. If one of the functions is put equal to zero one has
the pleasant property that the Einstein tensor of the averaged metric is 
identical to the averaged Einstein tensor.
 We also present a  simple covariant extension to static 
perturbations in four dimensions. Unfortunately the result for the 
Einstein tensor cannot be extended to the four dimensional case. 
 
\vfill \centerline{July 2009}

\end{titlepage}

\setcounter{equation}{0}\addtocounter{saveeqn}{1}%

\section{Introduction}

The energy-momentum-tensor used in
cosmological models is an average over the non homogeneous  
tensor present in nature. Therefore one needs an averaging prescription
for the energy momentum tensor and for the metric. This
provides a fundamental problem because of the freedom of choice of 
coordinates.
The averaging problem  in general relativity was first raised by 
Shirkov and Fisher \cite{ShirFi} in 1963.   The authors of \cite{ShirFi} 
suggested to
integrate the metric tensor over a four dimensional volume with the 
familiar factor $\sqrt{-g}$ in the measure. Such an expression is, 
however, not 
covariant due to the freedom of performing local transformations.
A covariant averaging prescription can be constructed by introducing a  
bivector $g_\alpha^{\beta }(x,x')$ of 
geodesic parallel displacement, as discussed 
in the appendix of \cite{Isaac}. This
transforms as a vector with respect to coordinate transformations at
either $x$ or $x'$ and maps a vector $A_\beta (x')$ to 
$\bar{A}_\alpha(x)= g_\alpha^{\beta }(x,x') A_\beta(x')$, analogously for 
higher order tensors.
An  averaging with the help of bivectors was also
used in the work of Zalaletdinov \cite{Zalat1} where the emphasis was on the 
commutativity of averaging and covariant differentiation. As remarked by 
Stoeger, Helmi, and 
Torres \cite{StoeHT}, the method of using a covariantly conserved 
bivector is not applicable to the metric, because 
the covariant derivative of the metric vanishes. The metric is therefore 
invariant under this averaging procedure.
In the  thesis of Behrend \cite{Behrend} the metric is represented by 
tetrads and the averaging performed over the latter. The tedrads are chosen 
according to a covariant minimalization prescription.

This is only a very brief survey of the literature. For more references,
as well as the implications for the fitting problem, back reaction, 
contributions to dark energy, we  refer e.g. to the monograph of 
Krasinski \cite{Kras} and
to the comprehensive recent reviews of Buchert \cite{Buchert} and of 
Malik and Wands \cite{MW}.

In a previous paper \cite{Gro} we gave an explicit solution. The derivation 
was rather complicated, furthermore we could not classify the complete
 set of solutions. The present approach is much more transparent
and straight forward. It naturally leads to a classification of all 
solutions  which fulfill the central requirements (1.1) and (1.2) below.

 Under a covariant averaging process 
we understand a prescription which has the following properties. 
Let two observers describe the same physics in different 
coordinate systems $S$ and $S'$, with metric tensors 
$g_{\mu\nu}$ and $g'_{\mu\nu}$. Both 
of them apply a definite averaging procedure in their respective systems,
resulting in the averaged metrics\newline  
$<g_{\mu\nu}>$ and $<g'_{\mu\nu}>$, respectively.
Then the results have to be connected by the same transformation as the 
original metric, i. e.

\begin{equation} 
<g'_{\mu\nu}> = <g_{\mu\nu}>'.
\end{equation}
In other words, the operations of averaging and of coordinate transformations 
have to commute. 

Furthermore, averaging over a region which is closely 
located around some point should, of course, reproduce the metric at this 
point. In this situation only the metric at the origin is relevant. This
means that a constant metric has to be reproduced by the averaging process, 
i. e.
 
\begin{equation} 
<\eta_{\mu\nu}> = \eta_{\mu\nu} \mbox{ for constant } \eta_{\mu\nu}.
\end{equation}
 It is rather obvious that space and time cannot be treated at the same 
footing in an
averaging prescription. Therefore, as usual, we have to assume that a 
reasonable foliation into space and time can be performed. 

In sect. 2 we present the most general lowest order three dimensional 
covariant averaging formula. 
The mapping of the metric $g_{kl}({\bf x}')$ to the averaged metric 
$<g_{mn}>({\bf x})$  is 
represented by a bitensor $K_{mn}^{kl}({\bf x}'-{\bf x})$ which we will specify
in detail. A product of 
bivectors, as frequently used in the literature, is not sufficient. 
The bitensor $K_{mn}^{kl}({\bf x}'-{\bf x})$
 is a superposition of a basis of six tensors, and 
contains three arbitrary functions $u(r)$, $v(r)$, and 
$w(r)$, which depend upon the distance $r = |{\bf x'} - {\bf x}|$ and
are only restricted by a prescribed behavior near the origin.
They are necessarily singular there. Our earlier formula in \cite{Gro} is
a special case of our general formula as it should be. 
In sect. 3 we show that, under the condition $w(r)=0$, we can use the 
same formula which works for the 
metric as well for the Einstein tensor. This implies that the Einstein 
equations for the averaged metric are identical to the averaged equations. 
In sect. 4 we show why iteration of the averaging formula does not make sense.
Therefore one has to choose reasonable functions from the beginning. 
In sect. 5 we discuss a simple four dimensional covariant 
generalization for 
static perturbations. The proof for the covariant averaging of the Einstein 
tensor ``almost'' goes through also in this case, but fails at the very end.

We claim that our formulae present the most general framework for a
lowest order averaging
prescription which respects (1.1) and (1.2).

\setcounter{equation}{0}\addtocounter{saveeqn}{1}%

\section{The most general lowest order three dimensional covariant
averaging formula}

Consider a perturbed flat metric $g_{kl}({\bf x}) = \delta_{kl} 
+ h_{kl}({\bf x})$.
In lowest order there must be a linear relation between the original metric
 and the averaged one. 
 For simplicity, we consider the averaging 
at the origin for the moment. 

\begin{equation}
<g_{mn}>(0) 
 =  \int K_{mn}^{kl}({\bf x})  \; g_{kl}({\bf x}) \frac{d^3x}{4\pi}.
\end{equation}
The same relation holds for the perturbation $h_{mn}$, because, as we shall 
see,
 averaging of the constant $\delta_{mn}$ gives back this constant as 
required by (1.2)
(the above $K_{mn}^{kl}({\bf x})$ is identical to the 
$\tilde{K}_{mn}^{kl}({\bf x})$ in \cite{Gro}).

In order to guarantee covariance with respect to rigid rotations, one has to
construct the most general structure of the tensor 
(tensor in the sense of linear algebra) $K_{mn}^{kl}({\bf x}) $ which is 
symmetric 
under the exchange
$m \leftrightarrow n$ and under $k \leftrightarrow l$. This is represented by 
a complete set of six independent tensors which are multiplied by
functions $A(r),\cdots , F(r),$ which depend only upon the distance 
$r=|{\bf x}|$:
\newpage
\begin{eqnarray}
K^{kl}_{mn}({\bf x}) & = & A(r) 
[\delta _m^k \delta _n^l 
+\delta _n^k \delta _m^l ] 
+  B(r) \delta _{mn} \delta ^{kl} 
+ C(r) \delta _{mn} \frac{x^k x^l}{r^2} 
+ D(r) \delta ^{kl} \frac{x_m x_n}{r^2} \nonumber\\
& &  
+ E(r) [\delta_m^k \frac{x_n x^l }{r^2}
+ \delta_m^l \frac{x_n x^k }{r^2}+ 
\delta_n^k \frac{x_m x^l }{r^2}
+ \delta_n^l \frac{x_m x^k }{r^2}] 
+ F(r) \frac{x_m x_n x^k x^l }{r^4}.
\end{eqnarray}
We emphasize that this is the most general tensor structure which one 
can write down.
Naive averaging would only make use of  $A(r)$, normalized to 
$\int _0^\infty 2 A(r) r^2 dr = 1$, while all the other 
functions $B(r), \cdots ,F(r)$ vanish.

Let us  now apply an arbitrary infinitesimal transformation 
$x^k = x'^k + \xi ^k$, which leads to a change 
$\delta g_{kl} = \xi_{k ,l} + \xi _{l ,k} \rightarrow 2\xi_{k ,l} $, 
when contracted with the symmetrical 
tensors in (2.2).  Covariance according to the requirement (1.1) implies that
the change of the rhs of (2.1) must be identical to the change of the lhs, 
i.e. to $\xi_{m ,n}(0) + \xi _{n ,m}(0)$. This means that the rhs cannot
depend upon $\xi_k({\bf x})$, except at the origin ${\bf x}=0$. To see  
how this 
can happen we perform a partial integration with respect to $x_l$. 
The change of the integrand on the rhs of (2.1) then becomes 
$-2 K^{kl}_{mn},_l \; \xi _k$.

This has to vanish for arbitrary $\xi _k$, which implies $K^{kl}_{mn},_l = 0$. 
Therefore $K^{kl}_{mn}$ has to be a curl with respect to $l$, and, by
symmetry, also with respect to $k$,
i.e. we can put

\begin{equation}
K^{kl}_{mn}({\bf x}) = \epsilon^k_{ab} \epsilon^l_{cd} \partial ^a \partial ^c 
T^{bd}_{mn}({\bf x}).
\end{equation}
The tensor $T^{bd}_{mn}({\bf x})$ has a decomposition analogous to (2.2), we 
denote the six radial functions by  the corresponding small letters 
$a(r),\cdots ,f(r)$.

\begin{eqnarray}
T^{bd}_{mn}({\bf x}) & = & a(r) 
[\delta _m^b \delta _n^d 
+\delta _n^b \delta _m^d ] 
+  b(r) \delta _{mn} \delta ^{bd} 
+ c(r) \delta _{mn} \frac{x^b x^d}{r^2} 
+ d(r) \delta ^{bd} \frac{x_m x_n}{r^2} \nonumber\\
& &  
+ e(r) [\delta_m^b \frac{x_n x^d }{r^2}
+ \delta_m^d \frac{x_n x^b }{r^2}+ 
\delta_n^b \frac{x_m x^d }{r^2}
+ \delta_n^d \frac{x_m x^b }{r^2}] 
+ f(r) \frac{x_m x_n x^b x^d }{r^4}.
\end{eqnarray}
This can be introduced into (2.3) and, after performing the differentiations, 
be compared with the general 
decomposition (2.2). The result is

\begin{eqnarray}
A(r) & = & -a''(r) - \frac{d(r)}{r^2} + 2 [\frac{e'(r)}{r} 
- \frac{e(r)}{r^2}] - \frac{f(r)}{r^2},
\nonumber\\
B(r) & = & 2 a''(r) + [b''(r)+ \frac{b'(r)}{r}]- \frac{c'(r)}{r} 
+ 2 \frac{d(r)}{r^2} 
- 4[\frac{e'(r)}{r}-\frac{e(r)}{r^2}] + 2 \frac{f(r)}{r^2},
\nonumber\\
C(r) & = & -2[a''(r) - \frac{a'(r)}{r}] - [b''(r) - \frac{b'(r)}{r}]
+[\frac{c'(r)}{r} - 2\frac{c(r)}{r^2}] + 4 [\frac{e'(r)}{r}-2\frac{e(r)}{r^2}]
- 2 \frac{f(r)}{r^2},
\nonumber\\
D(r) & = & -2[a''(r) - \frac{a'(r)}{r}] 
+[d''(r)  + \frac{d'(r)}{r} - 4 \frac{d(r)}{r^2}] 
+4 [\frac{e'(r)}{r}-2\frac{e(r)}{r^2}]
-  [\frac{f'(r)}{r}+4\frac{f(r)}{r^2}],
\nonumber\\
E(r) & = & [a''(r) - \frac{a'(r)}{r}] 
- [\frac{d'(r)}{r} - 2 \frac{d(r)}{r^2}]
- 2 [\frac{e'(r)}{r} - 2\frac{e(r)}{r^2}] + 2 \frac{f(r)}{r^2},
\nonumber\\
F(r) & = & [-d''(r) + 5\frac{d'(r)}{r} - 8 \frac{d(r)}{r^2}] 
+ [\frac{f'(r)}{r} - 4 \frac{f(r)}{r^2}]. 
\end{eqnarray}
Up to now we have six functions $a(r),\cdots ,f(r)$. But 
only three of them are relevant. To see this we first put  
 $a'(r) = \tilde{a}(r)/r$ and $b'(r) = \tilde{b}(r)/r$. One then finds that 
$e(r)$ and $\tilde{a}(r)$ only appear in the combination 
$e(r) - \tilde{a}(r)/2$,
while $c(r)$ and $\tilde{b}(r)$ only appear in the combination 
$c(r) - \tilde{b}(r)$. One thus can rename these combinations as $e(r)$ and 
$c(r)$, i. e. one can put $a(r) = b(r)=0$ in (2.5).
The next step is more subtle. Introduce two new functions $g(r)$ and $w(r)$ by
$g(r) = [c(r)+f(r)]/2$ and $w(r)/r^2 = C(r) - D(r)$. 
Eliminate 

$c(r) = [- r^2d''(r)/4 - r d'(r)/4  + d(r)] + [ rg'(r)/2 +g(r)] -w(r)/4$, and

$f(r) =  [r^2d''(r)/4 + r d'(r)/4  -d(r)] - [rg'(r)/2-g(r)] +w(r)/4$,
 
and use the functions $g(r)$ and $w(r)$ instead of $c(r)$ and $f(r)$. If we 
further put 
$g(r) = r\tilde{g}'(r)-2\tilde{g}(r)$, and eliminate $d(r)$ by introducing the 
 function $u(r) = d(r) - 2 \tilde{g}(r)$,
we  find that $e(r)$ and $\tilde{g}(r)$ only appear in the combination 
$v(r) = e(r) -\tilde{g}(r)$. Therefore we  end 
up with three relevant
functions $u(r),v(r),w(r)$.  (These $u(r),v(r)$ have, of course, 
nothing to do with the projective coordinates $u,v$ introduced in \cite{Gro}.) 
We thus have found the following representations for the functions 
$A(r), \cdots ,F(r)$ in (2.2):

\begin{eqnarray}
A(r) & = &  - \frac{1}{4}[u''(r)+ \frac{u'(r)}{r}] + 2 [\frac{v'(r)}{r} 
- \frac{v(r)}{r^2}] - \frac{1}{4}\frac{w(r)}{r^2},
\nonumber\\
B(r) & = & \frac{1}{4}[r u'''(r) +5 u''(r) -\frac{u'(r)}{r}] 
- 4[\frac{v'(r)}{r}-\frac{v(r)}{r^2}] 
+ \frac{1}{4} [\frac{w'(r)}{r} + 2 \frac{w(r)}{r^2}],
\nonumber\\
C(r) & = & -\frac{1}{4}[r u'''(r) +3 u''(r) -3 \frac{u'(r)}{r}] 
 + 4 [\frac{v'(r)}{r}-2\frac{v(r)}{r^2}] - \frac{1}{4} \frac{w'(r)}{r},
\nonumber\\
D(r) & = & -\frac{1}{4}[r u'''(r) +3 u''(r) -3 \frac{u'(r)}{r}] 
 + 4 [\frac{v'(r)}{r}-2\frac{v(r)}{r^2}] - \frac{1}{4} [\frac{w'(r)}{r} 
  + 4 \frac{w(r)}{r^2}] 
\nonumber\\
& = & C(r) - \frac{w(r)}{r^2},
\nonumber\\
E(r) & = & \frac{1}{2}[u''(r)- \frac{u'(r)}{r}]
- 2 [\frac{v'(r)}{r} - 2\frac{v(r)}{r^2}] + \frac{1}{2} \frac{w(r)}{r^2},
\nonumber\\
F(r) & = & \frac{1}{4}[r u'''(r) - 5 u''(r) + 13 \frac{u'(r)}{r} 
-16 \frac{u(r)}{r^2}] 
+ \frac{1}{4} [\frac{w'(r)}{r} -  4 \frac{w(r)}{r^2}]. 
\end{eqnarray}
We next demand that the averaging over a region  which is 
closely localized around the origin should give back the metric $g_{mn}(0)$ 
at the origin.
In this case we  may  put $g_{kl}({\bf x}) = g_{kl}(0)$ in (2.1) and take 
it out in 
front of the integral. The angular averages $\int d\Omega/4\pi$ can 
be performed using 
 
 \begin{equation}
\frac{x_m x_n}{r^2} \rightarrow \frac{1}{3}\delta_{mn}, \quad 
\frac{x_m x_n x^k x^l}{r^4} \rightarrow 
\frac{1}{15}(\delta_{mn} \delta^{kl} 
+ \delta_m^k \delta_n^l + \delta_n^k \delta_m^l)\quad \mbox{ etc.}
\end{equation}
Comparing the factors in front of the terms $g_{mn}(0)$ and 
$g_i^i(0)\delta_{mn}$ on both sides one 
thus obtains

\begin{equation}
1 = \int_0^\infty [2 A(r) + \frac{4}{3} E(r) + \frac{2}{15}F(r)]r^2 dr, \;
0 = \int_0^\infty [B(r) + \frac{1}{3}C(r) +  \frac{1}{3}D(r) 
+ \frac{1}{15}F(r)]r^2 dr.
\end{equation}
This derivation also shows the validity of (1.2), i. e. that the averaging 
of a constant gives back this 
constant. Therefore our averaging formula may be applied to the metric 
$g_{mn}$, as well as to the perturbation $h_{mn}$.
The relations can also be rephrased in the property

\begin{equation}
 \int K_{mn}^{kl}({\bf x}) \frac{d^3x}{4\pi} = 
\frac{1}{2} [\delta _m^k \delta _n^l + \delta _n^k \delta _m^l ]. 
\end{equation}
Inserting the representations (2.6) one finds that the  integrands in (2.8) can
be written as derivatives. Assuming that there are no boundary terms at 
infinity one thus obtains

\begin{eqnarray}
1 & = & [-\frac{1}{30} r^3 u''(r) + \frac{1}{10} r^2 u'(r) 
+ \frac{8}{15} r u(r) 
- \frac{4}{3} r v(r) - \frac{1}{30}r w(r)]_{r=0},\nonumber\\
0 & = & [-\frac{1}{10} r^3 u''(r) - \frac{11}{30} r^2u'(r) 
+ \frac{4}{15}r u(r) + \frac{4}{3}r v(r) - \frac{1}{10}r w(r)]_{r=0}.
\end{eqnarray}
The derivation above makes clear that the representation (2.1), (2.2), 
together with the form (2.6) and the boundary conditions (2.10),
is the most general first order covariant averaging formula in three 
dimensions.

Before we discuss the implications of (2.10) we consider the boundary terms 
at the origin which arise 
from the partial integration of $2 \xi_k,_l$, where $\xi_k$ was the shift of 
an infinitesimal transformation. Restricting the integration to the outside 
of a small sphere of radius $\epsilon$ these are just the surface terms which
arise from Gau{\ss}'s  theorem. Using $K^{kl}_{mn},_l = 0$, one has 

\begin{equation}
2 \int K_{mn}^{kl}({\bf x}) \xi_k,_l({\bf x}) \frac{d^3x}{4\pi}
= - 2 \int_{r=\epsilon} K_{mn}^{kl}({\bf x}) \frac{x_l}{\epsilon} 
\xi_k \epsilon^2\frac{d \Omega}{4\pi}.
\end{equation}
Because  invariance with respect to translations and to 
rigid rotations around the point of consideration is manifest, one can 
restrict to transformations which leave the origin fixed,  
 such that $\xi _k ({\bf x}) = \xi_{k,i} (0)x^i + O(\epsilon^2)$.
If we insert this into (2.11) and use the representation (2.2) for  
$K_{mn}^{kl}({\bf x})$ we can 
perform the angular averaging. The result has to be identical to the change 
$\xi_{m ,n}(0) + \xi _{n ,m}(0)$ of the lhs. Comparing the factors of
$\xi_{m ,n}(0) + \xi _{n ,m}(0)$ and of $2 \delta _{mn} \xi _k^k(0)$ 
on both sides one obtains

\begin{eqnarray}
1 & = & - \bigg[\frac{2r^3}{15} [5 A(r) + D(r) + 7 E(r) + F(r)]\bigg]_{r=0}, 
\nonumber\\
0 & = & - \bigg[ \frac{r^3}{15} 
[5B(r)+5C(r) +  D(r) + 2 E(r) + F(r)]\bigg] _{r=0}. 
\end{eqnarray}
If one introduces the representations (2.6) for $A(r), \cdots , F(r)$, one
obtains again the conditions (2.10).

The boundary conditions (2.10) have drastic consequences for 
 the behavior of the functions at the origin. They imply 

\begin{eqnarray}
& & u(r) = u_{[-1]}/r + O(1)  ,  v(r) = v_{[-1]}/r + O(1),
w(r) = w_{[-1]}/r + O(1), \mbox{ with }
\nonumber\\
& & u_{[-1]} =  \frac{5}{4} + \frac{1}{6} \; w_{[-1]} ,\quad v_{[-1]} 
= - \frac{13}{32} + \frac{1}{48} \; w_{[-1]}.
\end{eqnarray}
This result is unpleasant but unavoidable. The functions $u(r),v(r),w(r)$ are
singular and behave like $1/r$ near the origin. This implies that the
functions $A(r), \cdots ,F(r)$ go like $1/r^3$. At first sight this might 
look as if the integrand in (2.1) is not integrable at the origin. This is,
however, not the case. Expand $g_{kl}({\bf x}) = g_{kl}(0) + O(x)$. The 
constant term  $g_{kl}(0)$ can be taken out of the integral, 
 the angular averaging of the $1/r^3$-term vanishes.  This constant term has 
just been treated  in detail. The rest is  
of order $r/r^3$ and therefore integrable in three dimensions. 

Nevertheless one would have preferred functions  $A(r), \cdots ,F(r)$ in the 
averaging formula which are smooth at the origin. But covariance definitely 
prohibits such a smooth behavior. If one considers the change of the 
integral which
arises from an infinitesimal coordinate transformation 
$x^k = x'^k + \xi ^k$, the boundary terms 
of the partial integration  have to reproduce the  change of the metric at 
the origin, i.e. $\xi_{m ,n}(0) + \xi _{n ,m}(0)$. This enforces the 
singular behavior of the functions. Smooth functions could not produce 
boundary terms.

In \cite{Gro} we presented a special solution for a  covariant 
averaging procedure. It contained a function $f(r)$ 
(which has nothing to do with the $f(r)$ in (2.4)), normalized to 
$\int _0^\infty f(r) dr = 1$, as well as two integrals of $f(r)$, namely
$F(r) = - \int _r^\infty f(r')dr'$ and $
G(r) = -\int _r^\infty f(r')/r' dr'$. This solution must be a 
special case of our general formula. To demonstrate this we have to 
introduce another integral $H(r) = - \int _r^\infty f(r')/{r'}^3dr'$. Then we
obtain our old solution if we put

\begin{equation}
u(r) = - \frac{5}{4} \frac{F(r)}{r} + \frac{15}{8} G(r) 
- \frac{5}{8} r^2 H(r), \; 
v(r) = \frac{13}{32} \frac{F(r)}{r} - \frac{5}{16} r^2 H(r).
\end{equation}
When making this comparison one has to take take care of the correct factors 
$r^2$. In \cite{Gro} we used the 
integration element $dr d  \Omega/4\pi$ because we had to perform several 
partial integrations with respect to $r$ there, while in (2.1) we use 
$d^3x/4 \pi = r^2 dr d  \Omega/4\pi$.

\setcounter{equation}{0}\addtocounter{saveeqn}{1}%

\section{Covariant averaging of the Einstein tensor}

Besides the metric, the Einstein tensor is the most important object 
in general relativity because it enters, together with the energy momentum 
tensor, directly  the field equations.
It would be highly desirable if one could average the Einstein tensor in 
exactly the same way as the metric tensor, and if the averaged Einstein
tensor would be identical to the Einstein tensor derived from the averaged 
metric. The problem that the averaged equations are not identical with the 
equations for the averaged metric would then disappear. We repeat and extend 
some of the steps of \cite{Gro} in order to make the paper self contained.

  In first order of the perturbation the Einstein tensor becomes 

\begin{equation}
2G_{mn}= h^i_i,_{mn} + h_{mn},^i_i - h_{im},_n^i - h_{in},_m^i 
- h^i_i,^j_j \delta_{mn} + h_{ij},^{ij}\delta_{mn}.
\end{equation}
Indices are raised and lowered with $\delta _{ij}$ here, so their position 
is in fact irrelevant.
The averaging formula (2.1) 
is now used for an arbitrary point ${\bf x}$, the
integration variables are denoted by a prime, and the tensor  
$K^{kl}_{mn}({\bf x}'-{\bf x})$ depends on the difference ${\bf x}'-{\bf x}$. 
The distance $r'$ now means $r' = |{\bf x}'-{\bf x}|$. 
Due to the singular behavior of $K_{mn}^{kl}({\bf x}'-{\bf x})$ at 
zero distance one has to treat this region separately. Let us assume that 
we calculate the Einstein tensor near the origin, and expand
\begin{equation}
h_{ij}({\bf x}') = h_{ij}(0) + h_{ij},^a (0)x'_a 
+ \frac{1}{2} h_{ij},^{ab}(0) x'_a x'_b
+ \hat{h}_{ij}({\bf x}').
\end{equation}
The term $\hat{h}_{ij}({\bf x}')$ is of order ${x'}^3$, no problems with 
potentially divergent contributions or boundary terms from partial 
integrations can arise. We begin with this term.

The first possibility is to average the metric in the way described before.   
Subsequently one calculates the Einstein tensor from (3.1), using the 
averaged metric on the rhs.
The factor of $\hat{h}_{kl}({\bf x}')$ in the integrand then becomes

\begin{equation}
I_{mn}^{kl} = K_i^{ikl},_{mn} + K_{mn}^{kl},_i^i 
- K_{im}^{kl},_n^i - K_{in}^{kl},_m^i 
- K_i^{ikl},_j^j\delta _{mn} 
+ K_{ij}^{kl},^{ij} \delta_{mn}.
\end{equation}
The second possibility is to calculate the Einstein tensor $G_{kl}$ with
the old metric and then average it with our formulae in exactly the same way 
as we averaged the metric tensor. Shift the partial derivatives from the 
 perturbation $h$ to $K$, and rename dummy indices where necessary such
that $h_{kl}$ appears in all six terms. The factor of $\hat{h}_{kl}({\bf x}')$ 
in the integrand  now becomes

\begin{equation}
J_{mn}^{kl} = K_{mn}^{ij},_{ij}\delta^{kl} 
+ K_{mn}^{kl},_i^i 
- K_{mn}^{ki},^l_i - K_{mn}^{li},^k_i 
- K_{mni}^i,^j_j \delta^{kl} + K_{mni}^i,^{kl}.
\end{equation}
Consider the difference $\Delta _{mnkl}  =  I_{mnkl} - J_{mnkl}$. If one 
combines the terms appropriately one has 

\begin{eqnarray}
\Delta _{mnkl} 
& = & (K_{ikl}^i,_{mn} - K_{mni}^i,_{kl}) 
- (K_{imkl},_n^i - K_{mnki},_l^i) 
- (K_{inkl},_m^i - K_{mnli},_k^i ) \nonumber\\
& & - (K_{ikl}^i,_j^j\delta _{mn} - K_{mni}^i,^j_j \delta_{kl}) 
+ (K_{ijkl},^{ij} \delta_{mn} -  K_{mnij},^{ij}\delta_{kl} ).
\end{eqnarray}
The further investigation can be greatly simplified if one decomposes 

\begin{equation}
K_{mnkl} = K_{mnkl}^{[S]} + K_{mnkl}^{[A]},
\end{equation}
where $K_{mnkl}^{[S]}$ is symmetric against the exchange 
$(m,n) \leftrightarrow (k,l)$ and $K_{mnkl}^{[A]}$ antisymmetric. 
Obviously $K_{mnkl}^{[S]}$
consists of the terms with $A(r), B(r),E(r),F(r)$ in (2.2), together with
the symmetric combination $[C(r) + D(r)] [\delta _{mn} x_k x_l/r^2 
+  \delta _{kl} x_m x_n/r^2]/2$, while $K_{mnkl}^{[A]}$
consists of the antisymmetric combination 
$[C(r) - D(r)] [\delta _{mn} x_k x_l/r^2 
-  \delta _{kl} x_m x_n/r^2]/2$. From (3.5) it is seen that the symmetry 
relations in $\Delta _{mnkl}$ are just reversed with respect to $K _{mnkl}$, 
i.e. the symmetric part 
$\Delta _{mnkl}^{[S]}$ is obtained from $K_{mnkl}^{[A]}$, while the
antisymmetric part $\Delta _{mnkl}^{[A]}$ is obtained from $K_{mnkl}^{[S]}$.
Clearly $\Delta _{mnkl}$ has a decomposition analogous to $K_{mnkl}$, with
coefficients $\hat{A}(r), \cdots ,\hat{F}(r)$, say.

Let us first investigate the antisymmetric part $\Delta _{mnkl}^{[A]}$
which arises from $K_{mnkl}^{[S]}$. Knowing the structure 
$\Delta _{mnkl}^{[A]} = [\hat{C}(r) - \hat{D}(r)] \;
[\delta _{mn} x_k x_l/r^2 
-  \delta _{kl} x_m x_n/r^2]/2$,
we can  simplify the investigation by taking the trace $k=l$, thus
we only need to calculate \\
$\Delta _{mnk}^{[A]k} =  [\hat{C}(r) - \hat{D}(r)] 
[\delta _{mn} - 3 x_m x_n/r^2]/2.$
The result is

\begin{eqnarray}
\lefteqn{\frac{\hat{C}(r) - \hat{D}(r)}{2}  = } \nonumber\\ 
& & -[B''(r) + \frac{C''(r)+D''(r)}{2}] 
+ [B'(r)+F'(r)]\frac{1}{r} + [C(r)+D(r)+2F(r)]\frac{1}{r^2}
\nonumber\\ 
& = & w''(r)/2 r^2- w'(r)/r^3 - 2 w(r)/r^4. 
\end{eqnarray}
The functions $u(r)$ and $v(r)$ have dropped out completely, the expression 
vanishes if $w(r) = 0$. To see that this condition is also necessary 
we only need to consider, e.g.
the contractions $m=n,k=l$ of the symmetric part $\Delta _{mnkl}^{[S]}$, 
which gives 

\begin{equation}
\Delta _{mk}^{[S]mk} =
-2w''(r)/r^2 - 2 w'(r)/r^3 + 8 w(r)/r^4. 
\end{equation}
Both (3.7) and (3.8) have to vanish  and this is the case if and only if 
$w(r) = 0$. 
This is equivalent to the equation $C(r) = D(r)$, i.e. to the symmetry 
relation $K_{mnkl}^{[A]} = 0$.

We have to check that the independence of the order of averaging also holds 
for the first three terms in the decomposition (3.2). For the constant 
and linear terms $h_{ij}(0) + h_{ij},^a (0)x'_a$ this is trivial, they
do not contribute to the Einstein tensor, irrespective of the order of 
averaging. In the average of the quadratic term 
$ h_{ij},^{ab}(0) x'_a x'_b/2$, we substitute ${\bf x}'-{\bf x} = {\bf y}$, 
such that
$x'_a x'_b = x_a x_b + x_a y_b + y_a x_b + y_a y_b$. In the first term one
can take out $x_ax_b$ in front of the integral, the remaining integral is
given by (2.9), therefore $ h_{ij},^{ab}(0) x_a x_b/2$ 
is reproduced. The other three terms which are linear and constant with 
respect to ${\bf x}$  do not 
contribute to $G_{mn}$. If, alternatively, we first calculate $G_{mn}$, which 
is constant for this contribution, it is as well preserved by the averaging.
So we have seen by direct evaluation that the possibly dangerous low order
terms in (3.2) do not generate problems.

We finally have shown that the Einstein tensor of the averaged metric is 
identical to the averaged
Einstein tensor of the original metric if and only if $w(r) = 0$,
which is equivalent to $C(r) = D(r)$, i.e. the symmetry relation 
$K_{mnkl} = K_{klmn}$. In \cite{Gro}
we had assumed this symmetry in order to simplify the discussion, we now 
have shown that this condition is necessary. In the following we will
always assume $w(r) = 0$.

 An important feature for the understanding of this property are  
the symmetry relations shared by our 
averaging formula and by the Einstein tensor. This becomes clear if
one writes 

\begin{equation}
2G_{mn} = T_{mn}^{kl} h_{kl},
\end{equation}
with the operator

\begin{eqnarray}
 T_{mn}^{kl}  & = & \delta^{kl} \partial _m \partial _n 
+ \frac{1}{2} (\delta _m^k \delta _n^l + \delta _n^k \delta _m^l)  
\partial ^i \partial _i
-  \frac{1}{2} 
(\delta _m^k \partial _n \partial ^l  + \delta _m^l \partial _n \partial ^k
+ \delta _n^l \partial _m \partial ^k
+\delta _n^k \partial _m \partial ^l  ) 
\nonumber\\
& & -\delta ^{kl} \delta _{mn} \partial ^i \partial _i 
+ \delta _{mn} \partial ^k \partial ^l.
\end{eqnarray}
Both tensors, $K_{mnkl}$ as well as $T_{mnkl}$, are symmetric under
$m \leftrightarrow n$, under $k \leftrightarrow l$, and under 
$(m,n) \leftrightarrow (k,l)$. These symmetries were essential in order
to show the vanishing of  the difference (3.5). 

The result for the covariant averaging of the Einstein tensor is certainly 
not trivial. For the Ricci tensor, which does not fulfill the above 
symmetry properties, the relation is not valid.

\setcounter{equation}{0}\addtocounter{saveeqn}{1}%

\section{Iteration and stability}

The general behavior of iterations is most easily first studied in a 
simple one dimensional toy model. Consider an averaging formula

\begin{equation}
<g>(x) \equiv g_{[1]}(x) = \int K(x-x')g(x')dx',
\end{equation}
with $K(x)$ real and even, and normalized to $\int K(x)dx=1$. Because (4.1) 
is a convolution, it is convenient to work with the Fourier transforms 
$\tilde{g}(p) =  \int g(x)e^{-ipx}dx$, etc. This implies 
$\tilde{K}(0)=1$ and $\tilde{g}_{[1]}(p) = \tilde{K}(p)\tilde{g}(p)$.
The iteration of order $n$ of the  averaging procedure  becomes

\begin{equation}
\tilde{g}_{[n]}(p) = \tilde{K}^n(p)\tilde{g}(p).
\end{equation}
From this it is immediately clear that iteration does not make much sense. If 
$|\tilde{K}(p)| >1$  for some values of $p$,
the iteration will diverge.  If $|\tilde{K}(p)| <1$ the iteration 
will converge to 0 for these $p$. A stable averaging prescription 
$\tilde{K}(p)^2 = \tilde{K}(p)$ will be
obtained if and only if $\tilde{K}(p)$ only takes the values 0 or 1. Thus
consider a (finite or infinite) sequence $a_0=0 < b_0 <a_1 <b_1 < \cdots$, and
put $\tilde{K}(p) = \sum_n \Theta(a_n<|p|<b_n)$. This implies
the general form $K(x) = (1/\pi x) \sum_n (\sin b_nx - \sin a_n x)$ for a
 stable averaging function in this simple toy model.

One can start a similar investigation for our three dimensional 
averaging formula. 
It is easy to formulate the iterations in Fourier space, but it appears 
hard, probably 
impossible, to fulfill the conditions for a stable solution together with the 
representation (2.6) and the boundary conditions (2.13). The requirement 
of covariance prevents a stable averaging procedure.

\setcounter{equation}{0}\addtocounter{saveeqn}{1}%

\section{Static perturbations in Minkowski space}

Our extension to the four dimensional case is rather modest but practical. 
 We consider a Robertson Walker 
metric with $k=0$. A substitution $r = r'/a(t)$, with $a(t)$ the cosmic scale
factor, brings the line element into the form $ds^2 = dr^2 - dt^2 + \cdots$, 
where the corrections are small as long as the region of averaging is 
small compared to the Hubble length. Therefore we can use  the Minkowski metric
 (1,1,1,-1) as the unperturbed metric. We  further assume that the 
perturbation is approximately static.
To keep this situation, we restrict the admissible transformations to
 rigid translations, 
rigid spatial rotations, and infinitesimal transformations which keep the 
time unchanged. This means that $\xi^0 = 0$, and $\xi ^m$ is
independent of $t$. Furthermore we can drop all time derivatives in the 
perturbed metric.
Under these restrictions the perturbations $h_{00}$ and $h_{m0}$ become 
gauge invariant.

We average the perturbation with the following simple ansatz. 

\begin{eqnarray}
<h_{mn}>({\bf x}) & = & \int  K_{mn}^{kl}({\bf x}'-{\bf x})  
\; h_{kl}({\bf x}') 
  \frac{d^3 x'}{4\pi}, \\
<h_{00}>({\bf x}) & = & \int P({\bf x}'-{\bf x})
h_{00}({\bf x}')   \frac{d^3 x'}{4\pi},\\
<h_{m0}>({\bf x}) & = & \int Q({\bf x}'-{\bf x})
h_{m0}({\bf x}')   \frac{d^3 x'}{4\pi}.
\end{eqnarray}
Here  $P({\bf x}'-{\bf x})$ and $Q({\bf x}'-{\bf x})$ have to be  
rotation invariant and correctly normalized.
Equations (5.1) - (5.3) are the simplest generalization 
of our previous formula. For static perturbations they are covariant
in the sense of (1.1) with respect 
to static transformations.

Let us now consider the (lowest order) Einstein tensor which, in the static 
case, reads

\begin{eqnarray}
2 G_{mn} & = & 2 G_{mn}^{(s)} + h^0_0,_{mn} -  h^0_0,^i_i\delta_{mn} ,\\
2 G_{00} & = & h^i_i,^j_j - h_{ij},^{ij} \\
2 G_{m0} & = & h_{m0},^i_i - h_{i0},_m^i.
\end{eqnarray}
Here $G_{mn}^{(s)}$ is the spatial part of the Einstein tensor in (3.1). 
Under the assumptions above, the additional terms in $G_{mn}$,  as well as
 $G_{00}$ and $G_{m0}$ 
are invariant under infinitesimal static transformations.

We  investigate whether  the Einstein 
tensor of the averaged metric can be identical to the averaged Einstein tensor.
Although everything looks promising at the beginning, we will 
obtain a negative answer at the end. One may therefore skip the rest of this 
section.

As in the previous section we treat the constant, linear, and 
quadratic terms which were split off in (3.2) separately. Again for these
the  averaging is independent of the order in which it is performed. 
We start with $G_{mn}$. For $G_{mn}^{(s)}$ 
 we have shown in the previous 
section that the result is independent of the order of averaging. We can
restrict to the additional contributions in (5.4).

If we average the perturbation $h_0^0$ 
according to (5.2) and introduce into (5.4)  we obtain the integrand
$[P,_{mn} - P,_i^i \delta_{mn}]\hat{h}^0_0$.  
If, alternatively, we first calculate $2 G_{mn}$ in (5.4) with the old metric 
and then average it in the same way as (5.1), i.e. replace 
$h_{kl}$ by $G_{kl}$  there on the rhs, and
shift the partial derivatives from the perturbation to the multiplying 
functions, we obtain the integrand
$[K_{mn}^{kl},_{kl} - K_{mnk}^k,_l^l]\hat{h}^0_0$. 

This gives the condition

\begin{equation}
P,_{mn} - P,_i^i \delta_{mn} = K_{mn}^{kl},_{kl} - K_{mnk}^k,_l^l.
\end{equation}
We next apply the same procedure to $G_{00}$. If we average the 
perturbations $h_i^i$ and $h_{ij}$ 
according to (5.1) and introduce into (5.5)  we obtain the integrand
$[K_i^{ikl},_j^j - K_{ij}^{kl},^{ij}]\hat{h}_{kl}$.
If, alternatively, we first calculate $2 G_{00}$ in (5.5) with the old metric 
and then average 
it in the same way as (5.2), i.e. replace $h_{00}$ by $G_{00}$ there,
and shift the partial derivatives from the perturbation to the multiplying 
functions, we obtain the integrand
$[ P,_j^j \delta^{kl} - P,^{kl} ]\hat{h}_{kl}$.

The condition $K_i^{ikl},_j^j - K_{ij}^{kl},^{ij} 
= P,_j^j \delta^{kl} - P,^{kl} $ which now arises is identical to
  (5.7) if  one renames the dummy indices $k,l$, subsequently
replaces $m,n$ by $k,l$, and uses
the symmetry $K_{mnkl} = K_{klmn}$. We thus only need
to consider (5.7) in the following.

 An elementary calculation under proper consideration of the singular behavior
at the origin gives (of course we  demand $C(r) = D(r)$, 
i.e. $w(r) = 0$ in this section)

\begin{equation}
K_{mn}^{kl},_{kl} - K_{mnk}^k,_l^l = K_1(r) 
\frac{x_mx_n}{r^2} + K_2(r) \delta_{mn} 
- \frac{4\pi}{2}[\partial _m \partial _n - \delta _{mn} \Delta] 
\delta^{(3)}({\bf r}),
\end{equation}
with 

\begin{eqnarray}
K_1(r) & = & 2[A''(r) - \frac{A'(r)}{r}] 
- 2[C''(r) + 2 \frac{C'(r)}{r} - 6 \frac{C(r)}{r^2}] 
- 4 [\frac{E'(r)}{r} - 2 \frac{E(r)}{r^2}], 
\nonumber\\
& & + 2 [\frac{F'(r)}{r} + 4 \frac{F(r)}{r^2}] 
\nonumber\\
K_2(r) & = & - 2[A''(r) + \frac{A'(r)}{r}] - 2[B''(r)+ 2\frac{B'(r)}{r}]
+ 2[\frac{C'(r)}{r} - \frac{C(r)}{r^2}] + 4 \frac{E'(r)}{r} 
\nonumber\\
& & - 2 \frac{F(r)}{r^2}.
\end{eqnarray}
The term with the $\delta$-function is most conveniently obtained by 
multiplying (5.8) with $x^ix^j$ and integrating. 
If one introduces the representation (2.6) and considers the behavior (2.13)
near $r=0$, one finds that the singular terms in $u(r)$ and $v(r)$ cancel,
$K_1(r)$ and $K_2(r)$ behave at most like $1/r^4$ for small $r$. By a suitable 
choice of the remaining freedom in $u(r)$ and $v(r)$ one could also remove any
singularities, but  the convergence of the integrals used in the following
is guaranteed anyhow. 
If one equates the factors
of $-[x_mx_n/r^2+\delta_{mn}]/r$, of $[x_mx_n/r^2-\delta_{mn}]$, and of 
$[\partial _m \partial _n - \delta _{mn} \Delta] 
\delta^{(3)}({\bf r})$ in  (5.7) one obtains 

\begin{equation}
P({\bf x}) = \hat{P}(r) - \frac{4\pi}{2} \delta^{(3)}({\bf r}),
\end{equation}
with

\begin{equation}
 \hat{P}'(r) = -\frac{r}{2}[K_1(r)+K_2(r)],\;
  \hat{P}''(r)  = \frac{1}{2}[K_1(r) -K_2(r)].
\end{equation}
This implies  the integrability condition

\begin{equation}
K_1'(r) + K_2'(r) + \frac{2 K_1(r)}{r} = 0.
\end{equation}
It is fulfilled if one introduces the representations (5.9) and (2.6).
The normalization of $\hat{P}$  becomes 
 
\begin{equation}
 \int _0^\infty r^2 \hat{P}(r) dr =
- \frac{1}{3} \int_0^\infty r^3 \hat{P}'(r)dr 
= \frac{1}{6} \int _0^\infty r^4 [K_1(r) + K_2(r)]dr =  \frac{3}{2}.
\end{equation}
From (5.10) this implies the correct normalization of $P$.
The last integrand in (5.13) (as well as the integrand with $K_1(r)$ or 
$K_2(r)$ alone) turns out to be a derivative when one introduces (5.9) 
and (2.6). 
Therefore no freedom is left, the 
integral is determined by the boundary term at zero, and the latter is fixed by
the behavior of $u(r)$ and $v(r)$ in (2.13).

This result is unpleasant. It fixes the integrals over $r^4 K_1(r)$ and 
$r^4 K_2(r)$, and leads to the unwanted $\delta$-function contribution in (5.8)
and (5.10). The $\delta $-function in (5.10) would imply that $<h_0^0>(x)$
would contain a contribution $-h_0^0(x)/2$, i. e. a contribution which 
is not averaged, and furthermore, with opposite sign. Such a 
contribution  cannot be tolerated. 

One could use a more general ansatz in (5.1), (5.2), where  terms
with $\delta_{kl} h^0_0$ and $(x_kx_l/r^2) h^0_0$ are inserted into the 
rhs of (5.1), and a 
gauge invariant combination of $h^k_k$  and $(x^kx^l/r^2) h_{kl}$
 into (5.2). We found that this does not help to solve the problem. 
At the end the relevant integrands  again turn out to be derivatives, 
and everything is fixed by the boundary conditions. 
We conclude that  the previous result about the covariant averaging of the 
Einstein tensor cannot be extended to four dimensions. 

Therefore we keep 
things simple and work with (5.1) - (5.3).
We may choose rather arbitrary functions $P$ and $Q$ in (5.2), (5.3),
of course with the correct normalization and somehow related to 
 the functions in (2.2). Nevertheless we favor the choice $w(r)=0$
also in the four dimensional case.

\setcounter{equation}{0}\addtocounter{saveeqn}{1}%

\section{Outlook and conclusions}

In this paper we presented a covariant averaging prescription which fulfills
two essential requirements. 

\begin{itemize}
\item Under coordinate transformations the averaged metrics are 
connected by the same transformation as the original metrics, i.e. 
$<g'_{\mu\nu}> = <g_{\mu\nu}>'$ (eq. (1.1)).
\item The averaging of a constant metric  reproduces this  metric, i.e.
$<\eta_{\mu\nu}> = \eta_{\mu\nu}$ for constant $ \eta_{\mu\nu}$ (eq. (1.2)).
\end{itemize}
In three dimensions we gave the complete solution of the problem. There
is a linear connection between the original perturbation 
$h_{kl}({\bf x}')$ and the averaged $<h_{mn}>({\bf x})$, represented
by a tensor (tensor in the sense of linear algebra) 
$K_{mn}^{kl}({\bf x'} - {\bf x})$. (Such a connection has also 
been discussed by Boersma \cite{Boersma}, although without going into details.)
The tensor $K_{mn}^{kl}({\bf x'} - {\bf x})$ is a superposition of a 
basis of six bitensors which are symmetric 
with respect to $m \leftrightarrow n$ and to $k \leftrightarrow l$. 
A product of bivectors, as sometimes suggested in the literature, 
is not sufficient.
The representation  contains three functions $u(r), v(r), w(r)$ which
depend upon the distance $r= |{\bf x}' - {\bf x}|$. They have to be singular 
in a definite way 
at the origin, in order to fulfill the requirement (1.1) of covariance.
For $w(r)=0$ we found the welcome property that the Einstein tensor can be 
averaged in the same way as the metric, and that the Einstein tensor 
belonging to the averaged 
metric is identical to the averaged Einstein tensor.

We have further seen  that it does not make sense to iterate the 
averaging procedure. Therefore one has to choose a 
reasonable  ansatz for the functions 
$u(r)$ and $v(r)$  from the beginning, of course with the correct 
boundary conditions. Three simple choices suggest themselves:

Exponential function:

\begin{equation}
u_{exp}(r) = \frac{5}{4} \frac{e^{-r/r_0}}{r}, \; 
v_{exp}(r) = - \frac{13}{32 } \frac{e^{-r/r_0}}{r}.
\end{equation}

Gaussian:

\begin{equation}
u_{gauss}(r) = \frac{5}{4} \frac{e^{-(r/r_0)^2}}{r}, \; 
v_{gauss}(r) = - \frac{13}{32 } \frac{e^{-(r/r_0)^2}}{r}.
\end{equation}

Averaging over a sphere:

\begin{equation}
u_{sphere}(r) = \frac{5}{4} \frac{(r_0-r)^4}{r r_0^4}\; \Theta(r_0-r), \; 
v_{sphere}(r) = - \frac{13}{32 } \frac{(r_0-r)^2}{r r_0^2}\; \Theta(r_0-r).
\end{equation}
In the last case we took care not to get $\delta-$ like contributions in
the derivatives from the 
boundary at $r=r_0$.

The functions $A(r), \cdots , F(r)$ can be easily obtained from this with 
the help of (2.6), there is no need to show the explicit expressions here.

The generalization to static perturbations in Minkowski space, for
many applications an excellent approximation to the realistic case, 
is quite simple. 
The property for the covariant averaging of the Einstein tensor could, 
however, not be generalized to the four dimensional case.

One can hope to obtain a sufficiently smooth metric after performing the 
average. But it is important to note that we
 are still free to perform gauge transformations. By an
unfavorable choice of gauge the ``smoothed'' metric can become wavy and 
irregular. All one can achieve is that the final metric becomes equivalent 
to a smooth metric.

The present investigation was already quite elaborate, it gave the 
mathematical framework for a covariant averaging prescription.  
Applications have to be postponed to forthcoming work. The central question 
to be investigated is,
 of course,  how far the averaging of inhomogeneities 
can mimic the presence of dark energy.
\\

{\bf Acknowledgement: } I thank Juliane Behrend for valuable discussions 
and for her interest in this work.

\end{document}